\documentclass[11pt]{article}
\pdfoutput=1

\newcommand{\Ro}{\ensuremath{{\cal R}_0}}

\def\d{{\rm d}}

\usepackage{amssymb}
\usepackage{authblk}
\usepackage{geometry}                		
\usepackage{graphicx}				

\usepackage{amsmath,amssymb}

\usepackage{pdfpages}
\usepackage{dsfont}
\title{Stochasticity and the limits to confidence when estimating ${\cal{R}}_0$ of Ebola and other emerging infectious diseases}

\author[1,4]{Bradford P. Taylor}
\author[2]{Jonathan Dushoff}
\author[1,3]{Joshua S. Weitz}
\affil[1]{School of Physics, Georgia Institute of Technology, Atlanta, GA, USA}
\affil[2]{Department of Biology and Institute for Infectious Disease Research, McMaster University, Hamilton, Canada}
\affil[3]{School of Biology, Georgia Institute of Technology, Atlanta, GA, USA}
\affil[4]{Corresponding author: bradfordptaylor@gmail.com}

\date{\today}

\begin{document}

\maketitle

\begin{abstract}
Dynamic models - often deterministic in nature - were used to estimate the basic reproductive number, ${\cal{R}}_0$, of the 2014-5 Ebola virus disease (EVD) epidemic outbreak in West Africa. Estimates of ${\cal{R}}_0$ were then used to project the likelihood for large outbreak sizes, e.g., exceeding hundreds of thousands of cases. Yet fitting deterministic models can lead to over-confidence in the confidence intervals of the fitted ${\cal{R}}_0$, and, in turn, the type and scope of necessary interventions. In this manuscript we propose a hybrid stochastic-deterministic method to estimate ${\cal{R}}_0$ and associated confidence intervals (CIs). The core idea is that stochastic realizations of an underlying deterministic model can be used to evaluate the compatibility of candidate values of ${\cal{R}}_0$ with observed epidemic curves.  The compatibility is based on comparing the distribution of expected epidemic growth rates with the observed epidemic growth rate given ``process noise'', i.e., arising due to stochastic transmission, recovery and death events. By applying our method to reported EVD case counts from Guinea, Liberia and Sierra Leone, we show that prior estimates of ${\cal{R}}_0$ based on deterministic fits appear to be more confident than analysis of stochastic trajectories suggests should be possible. Moving forward, we recommend including a hybrid stochastic-deterministic fitting procedure when quantifying the full ${\cal{R}}_0$ CI at the onset of an epidemic due to multiple sources of noise. 

\end{abstract}

\section{Introduction}
%previous models
The SEIRD model of Ebola virus disease (EVD) dynamics, 
introduced by Legrand and colleagues~\cite{legrand2007understanding},
considers the transitions among susceptible, exposed, infectious, recovered and dead (but unburied) individuals. Variants of this core model have been
utilized to estimate the
basic reproductive number, ${\cal{R}}_0$, of EVD
from incidence and cumulative case 
data in the 2014-5 epidemic in West Africa (e.g., ~\cite{althaus2014,Gomes2014,Fisman2014,teamebola,nishiura2014,pandey_2014,lewnard_2014,meltzer_2014,rivers_2014}).
In some instances these estimates include 95\%
confidence intervals. For example, one  of
the first mathematical epidemiology papers published
in response to the ongoing epidemic in W. Africa
estimated the ${\cal{R}}_0$ of EVD to be:
1.51 (95\% CI 1.50--1.52) in Guinea;
1.59 (1.57--1.60) in Liberia; 
and 2.53 (2.41--2.67) in Sierra Leone~\cite{althaus2014}.
The confidence limits appear over-confident, particularly for
Guinea and Liberia.

Subsequently, the World Health Organization (WHO) Ebola Response Team 
analyzed a model with
pre- and post-death transmission and estimated ${\cal{R}}_0$ to be:
1.71 (1.44--2.01) for Guinea;
1.83 (1.72--1.94) for Liberia; 
and 2.02 (1.79--2.26) for Sierra Leone~\cite{teamebola}.
Similarly, a follow-up model for EVD in the Montserrado region of
Liberia estimated ${\cal{R}}_0$ to be
2.49 (2.38--2.60)~\cite{lewnard_2014}.
Comparison of these case studies shows that models
can provide incompatible inferences with non-overlapping CIs despite similar
data.  Obviously, differences will arise in model structure,
data range and quality, and the treatment of noise.  Yet, are even
these more recent studies over-confident about the precision of ${\cal{R}}_0$ estimates for EVD?

Aaron King and colleagues posed a similar question 
and cautioned that reported confidence intervals (CIs) of ${\cal{R}}_0$ for EVD
are likely too narrow whenever they neglect stochasticity in the disease 
transmission process~\cite{king_2015}. 
Such over-confidence is further heightened by fitting deterministic models 
to cumulative case curves (CCCs).  A CCC is a monotonically increasing function
of time, representing the total number of individuals reported to have
become infected during an outbreak.  Individual time points within CCCs are not independent, and so ``error'' in fitting deterministic models
can appear artefactually low.  Such low error in fits -- if not properly
accounted for -- can lead to a misleading interpretation of
overly narrow CIs for ${\cal{R}}_0$.
Instead King et al.~\cite{king_2015} recommend a stochastic-based fitting procedure to incidence
case curves (ICCs), to account for observation noise and process noise
into estimates of ${\cal{R}}_0$ and its associated CIs.

Here, we propose a hybrid stochastic-deterministic
approach to address similar issues. 
The key differentiating theoretical feature of our approach from
that proposed by King et al.~\cite{king_2015}, is that we 
address uncertainty in generation-interval distributions
and its effect on ${\cal{R}}_0$. The generation-interval distribution, $g(a)$,
is the normalized fraction of secondary cases caused by an infectious individual
at ``age'' $a$ since infection. In the case of EVD, there
was -- and is -- uncertainty with respect to  transmission event times,
including that of post-death transmission~\cite{weitz_2015, nielsen_2015}.

We apply our hybrid approach retrospectively to estimate 
the CIs for ${\cal{R}}_0$ for EVD in the summer of 2014, coincident
with the release of the first projections for the potential
size of the outbreak.
We conclude that many early estimates of the CIs for ${\cal{R}}_0$
for EVD were almost certainly over-confident. The over-confidence
arose, at least in part, because estimates did not account for the effects of process noise and for uncertainty in disease generation times. More generally, we explain how
models of disease transmission can
be adapted to our approach -- by using a principled filtering
method to identify ensembles of simulated stochastic trajectories
``compatible'' with a single, measured epidemic outbreak. Overall, we propose a flexible framework that is easily adaptable to the specific dynamics of a particular emerging epidemic while remaining simple to computationally implement.

\section{Methods}
\subsection{Measuring Compatible Epidemic Trajectories given Stochasticity}

Dynamic models provide a means to estimate the overall intensity of an outbreak from measurements of infectious case data.  The intensity is usually quantified in terms of the basic reproduction number, \Ro ,--the total number of secondary cases caused by a single infectious individual in an otherwise naive population. An alternative metric of intensity is the growth rate of disease incidence, $r_0$, which is the reciprocal of the characteristic time of exponential growth, $\tau_c$. These quantities are closely related to the doubling time $\tau_2 = \tau_c\ln{2} = \frac{\ln{2}}{r_0}$. These two metrics of disease outbreak intensity, ${\cal{R}}_0$ and $r_0$, are related but not equivalent; they are linked by the generation-interval distribution, $g(a)$.
For example, in a simple SIR model then $g(a)=\frac{1}{T}e^{-a/T}$ where
the average infectious period is $T$.  As is well known,
the theoretically expected epidemic growth rate is $r_0=({\cal{R}}_0-1)/T$.
If the generation-interval distribution is known, an estimate of 
${r}_0$ implies a \emph{unique} corresponding estimate of \Ro, e.g., it is
${\cal{R}}_0=1+r_0T$ in the case of a SIR model. 
Conversely, when the generation-interval distribution is uncertain, then a range of values of ${\cal{R}}_0$ may be compatible with a given rate of increase in disease incidence -- in particular, an increase in the estimated mean generation-interval $T$ would lead to an increase in estimated \Ro.
This identifiability problem linking the measured
value of ${r}_0$ and the estimated value of ${\cal{R}}_0$ hampers efforts to estimate the potential scope of the disease outbreak over the long-term~\cite{weitz_2015,eisenberg2015}. 

Uncertainty in generation intervals is one of many challenges in estimation of ${\cal{R}}_0$.
Other important sources of uncertainty include the intrinsic stochasticity of the epidemic, the case-reporting process and uncertainty in the structural mode of disease transmission.   
Ignoring or under-estimating uncertainty can lead to over-confident estimates of ${\cal{R}}_0$.  
The premise of our approach is that a measured time series of infectious
case data represents a single trajectory from an ensemble
of stochastic trajectories given a set of underlying and unknown parameters.
The observed trajectory of disease incidence need not grow exponentially at theoretically expected values. Rather there are many trajectories that could
appear statistically indistinguishable from the observed epidemic outbreak. We term these trajectories: ``\textbf{Com}patible \textbf{E}pidemic \textbf{T}rajectories given \textbf{S}tochasticity'' or COMETS.
The expected variation in the growth rates of COMETS is quantifiable and can be used to bound the confidence of an ${\cal{R}}_0$ estimate from a single stochastic trajectory.

The hybrid stochastic-deterministic approach to estimate
the CIs of ${\cal{R}}_0$ from synthetic data relies on an inverse problem approach to regression-based fits. 
The approach involves the following steps:
\begin{itemize}
\item A range of deterministic models is considered that vary 
in disease-associated parameters, including ${\cal{R}}_0$. 
\item For each model and
fixed parameter set within the range, we simulate an ensemble of stochastic trajectories and utilize a metric to compare the simulated trajectories to the case data. The metric is the regression-based
estimate of the characteristic time, $\hat{\tau}_c$, given either
a CCC or an Incidence Case Curve (ICC).
\item A value of ${\cal{R}}_0$ is defined to be compatible with a data set if the value of ${\tau}_c$ inferred from the data, $\hat{\tau}_c$,
lies in the middle portion of the distribution of values of $\tau_c$ inferred from the ensemble of simulations associated with that value, $\tilde{\tau}_c$. 
Throughout this paper, we focus on the middle $95\%$ of the distribution, and use the symbols $\mbox{}~\hat{\mbox{}}$ and $\mbox{}~\tilde{\mbox{}}$ to denote estimates from empirical data and simulations, respectively.

\item We estimate CIs associated with ${\cal{R}}_0$ by identifying the range of deterministic models that can yield dynamics compatible with the case data, i.e., by identifying COMETS.
\end{itemize}

This series of steps can be adopted to any epidemic context. Moreover,
the well-known problems with inferring CIs for ${\cal{R}}_0$
based on the quality of model
fits to a single CCC do not arise via the COMETS approach.
In the next section we illustrate why the method is relevant
to EVD by highlighting the extent to which process noise drives
uncertainty in the distribution of $\tilde{\tau}_c$ of stochastic epidemic trajectories in a SEIRD framework.

\subsection{Stochastic trajectories of epidemic outbreaks have
substantial variation in epidemic growth rates}
\label{sec.variation}
Stochastic variation in the timing of discrete
transmission events can generate variation in estimates
of the realized growth rate of an epidemic, $r_0$. 
The amount of variation depends on the particular disease model, disease parameters, and the time-scale over which $r_0$ is estimated.
As an example inspired by EVD, we consider disease dynamics based on a SEIRD model:
\begin{eqnarray}
\frac{d S}{d t} &=& -\beta_I S I/N -\beta_D S D/N,  \\
\frac{d E}{d t} &=& \beta_I S I/N +\beta_D S D/N -  E/T_E, \\
\frac{d I}{d t} &=& E/T_E - I/T_I, \\
\frac{d R}{d t} &=& (1-f) I/T_I,  \\
\frac{d D}{d t} &=& f I/T_I - D/T_D. \label{eq.D.dynamics}
\end{eqnarray}
We assume that the average latent period is $T_E=11$ days, the average infectious period is $T_I=6$ days, there is a $f=.7$ chance of an infected individual ultimately dying and there is an average time of $T_D=4$ days before burial. The total population is set at $N=10^6$.
We set the disease transmission rates to be $\beta_D=0.20$ and
$\beta_I=0.25$, meaning that
a fraction $\rho_D=\frac{{\cal{R}}_0(\mathrm{dead})}{{\cal{R}}_0}=0.25$
of transmission is attributable to post-death transmission (see \ref{sec.app.generate}). 
In a deterministic framework, epidemics that obey such a model given those parameters should increase 
with a characteristic time of $\tau_c=21$ days~\cite{wallinga_2007,weitz_2015}. Note, that our choice of models here and throughout the paper inherently alter the estimates of the confidence intervals. Thus we caution that identifying an appropriate underlying model structure is a key but separate issue to the analysis process.

We simulate an ensemble of $10^4$ stochastic realizations of this SEIRD model 
beginning with a single infectious individual
(see \ref{sec.app.model}).
We consider only realizations that produce at least 50 total cases (approximately 58\% of all realizations).
This threshold acts as a trigger, from which point we track the time series of incidence, accumulating cases at exact daily intervals.
We do not account for reporting error in this example.
We estimate the growth rate by fitting an incidence curve (CCC or ICC) from the trigger time $t_0$ until $t_f=t_0+2\tau_c$. For example, in the case of $\tau_c=21$ then the fitting period is 42 days in duration.
The measured epidemic growth rate, $\hat{r}_0$, for a given trajectory is the slope of the best-fit line to log-transformed censused time series assuming errors are Poisson distributed. Phenomenologically these errors are attributed to observation noise. Thus our presented model only implicitly includes observation noise. That is we do not introduce external noise to our data representing errors in reporting. Note that the results depend on the choices for $t_0$ and $\tau_c$. Hence, the modeler should choose $t_0$ such that the error in the fitting line to the case data is satisfactoraly small. Meanwhile for an emerging epidemic, $\tau_c$ should be dictated by the total time elapsed from the onset. The demonstration of our methodology does not rely on our specific choices in this paper.

\begin{figure}[h]
\includegraphics[width=0.95\textwidth]{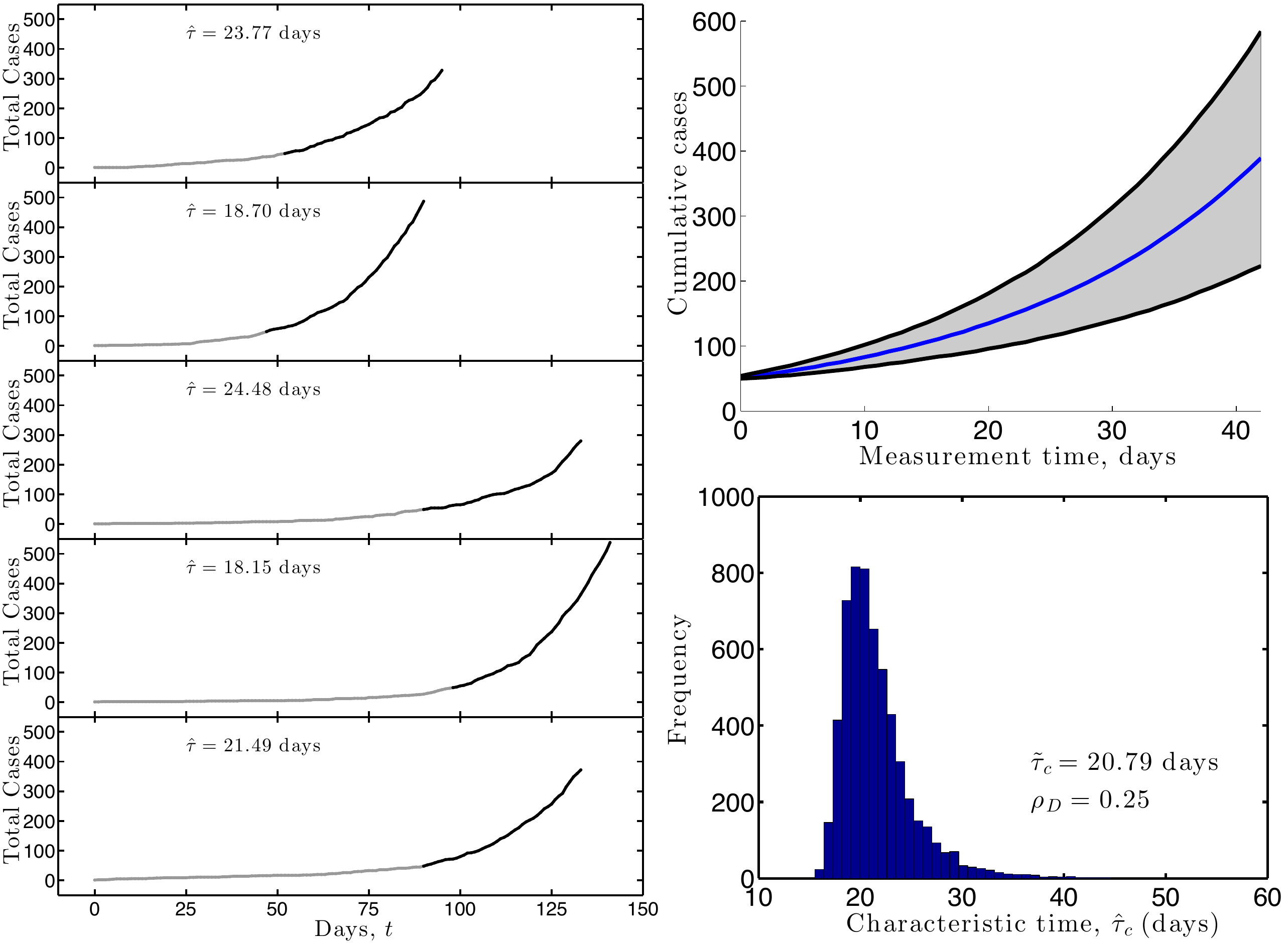}
\caption{{\bf Stochastic realizations of a SEIRD epidemic include substantial
variability in epidemic growth rate.}
(Left) Each panel denotes a randomly chosen trajectory for which the cumulative case count exceeded 50 (gray period) followed by a 42 day measurement period
(black period).  The estimated $\tilde{\tau}_c$ in the measurement period
is denoted in the upper left of each panel.
(Top Right) Variation in cumulative case counts once a threshold is
reached.  The results are for $5817$ epidemics where the measurement
time denotes the period after the first day the trigger was crossed.
The solid blue line denotes the median number of cases and the shaded region the central 95\% of simulations.
(Bottom Right) Variation in the characteristic time estimated from simulation data, $\tilde{\tau}_c$.  
In all cases, simulations correspond to stochastic simulations of
a SEIRD model in which $N=10^6$, $\beta_I=0.25$, $\beta_D=0.2$, 
$\sigma=1/11$, $\gamma=1/6$, $f=0.7$ and $\rho_D=0.25$.
The theoretically expected characteristic time is $\tau_c=21$ days. The median characteristic time is $\tilde{\tau}_c=20.73$ days.}
\label{fig.traject}
\end{figure}

The left panel of Figure~\ref{fig.traject} shows five examples of stochastic epidemic trajectories. 
The time from the index case until the point at which the epidemic ``takes off'' (i.e., reaches a threshold number of cases), is highly variable.
The upper right panel of Figure~\ref{fig.traject} shows that, even after take-off, there is substantial variation in growth rate between different epidemic realizations.
The trajectories have been time shifted so that $t=0$ is now defined as the day where the trigger population is reached or surpassed. 
There is over 40\% variation above and below the number of infectious cases at $t=42$ days.
The bottom right panel of Figure~\ref{fig.traject} shows the variation in characteristic times estimated from realized data,
within an ensemble of epidemic trajectories.  
The median characteristic time is 20.73 days with a CI of $[17.0,30.1]$ days.  
In the appendix, we perform a sensitivity analysis of our method by considering different trigger thresholds. There, the distribution of the characteristic time widens as the cumulative trigger population decreases (see Figure \ref{fig.Imin_compare}).
This exploratory analysis reveals substantial variability in the growth rate of epidemic trajectories.  
Quantifying the uncertainty in the characteristic time of epidemic outbreaks in an ensemble is the basis for estimating the CI for ${\cal{R}}_0$.

\subsection{The expected uncertainty in ${\cal{R}}_0$ for
an EVD-like outbreak
as estimated from a single stochastic trajectory}
\label{sec.single}

\begin{figure}[h]
\includegraphics[width=0.475\textwidth]{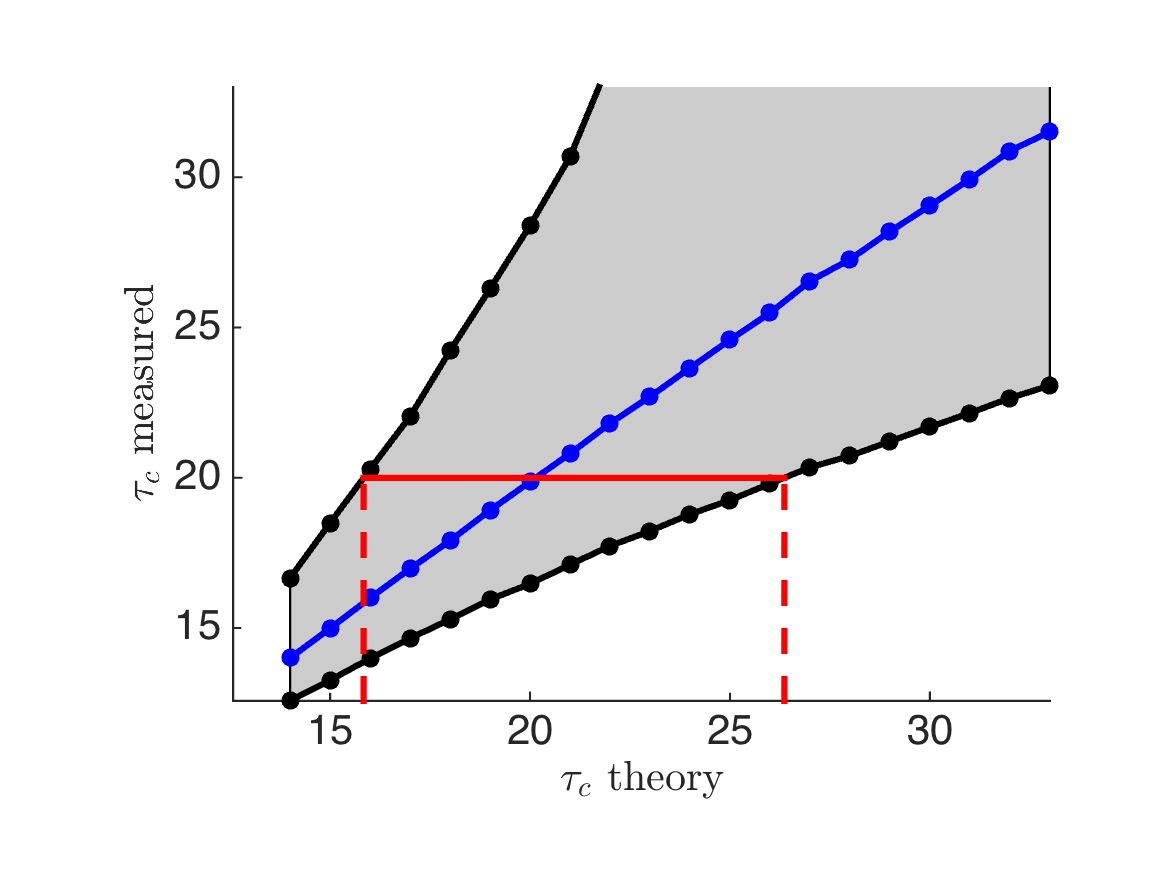}
\includegraphics[width=0.475\textwidth]{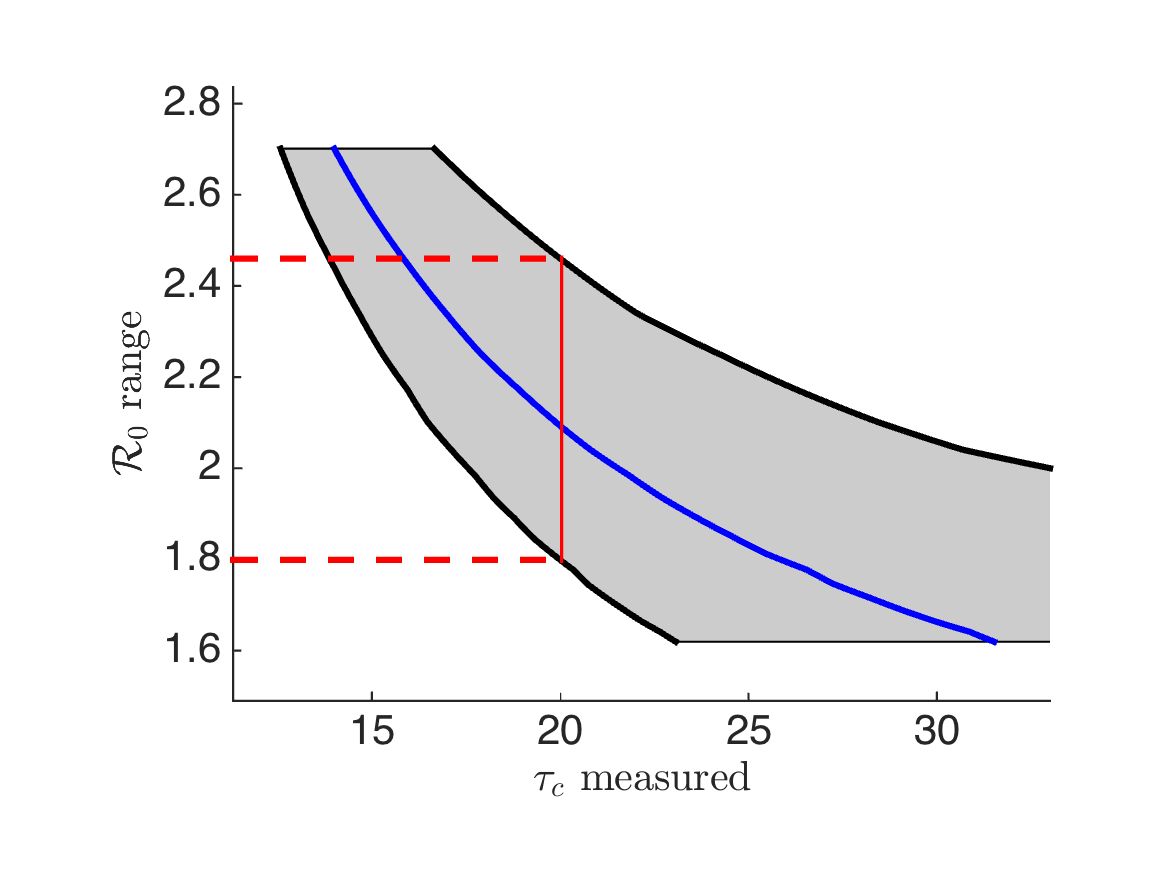}\\
\caption{{\bf Fundamental limits on inferring ${\cal{R}}_0$ from
a single stochastic epidemic trajectory.} Variation in the CIs of theoretical (left) $\tau_c$ and corresponding (right) ${\cal{R}}_0$ when using CCCs.  
See text for details on methods and interpretation.}
\label{fig.R0_var}
\end{figure}

Here, we estimate the uncertainty in ${\cal{R}}_0$ for an EVD-like outbreak
with SEIRD dynamics.  As in the previous section, 
the fraction of post-death transmission 
is assumed to be $\rho_D=0.25$.
Then, $\beta_I$ and $\beta_D$ are varied to yield different expected
characteristic times for the case counts, $\tau_c$ (see analytic relationships in \ref{sec.app.generate}). 
For each parameter set, we simulate $10^4$ stochastic trajectories with the same censusing conditions as in the previous section, conditioned on the fact that the epidemic
continues throughout the sampling period. 
Variation in the measured $\tilde{\tau}_c$ given a range of theoretically
expected values of $\tau_c$ 
is shown in the left panel of Figure~\ref{fig.R0_var}. This figure is the
basis for identifying COMETS and, in turn, for estimating the CIs for
${\cal{R}}_0$ at the onset of the epidemic.

The conventional way to interpret Figure~\ref{fig.R0_var} is
as a ``forward problem'' -- as introduced in the previous section.  
In the forward problem approach,
Figure~\ref{fig.R0_var} depicts variation in the
measured characteristic  time of stochastic epidemic trajectories,
$\tilde{\tau}_c$, given variation in
the theoretically expected $\tau_c$.  The variation
is summarized as a probability distribution
$p(\tilde{\tau}_c|\tau_c)$. The central 95\% of the distribution of
$p(\tilde{\tau}_c|\tau_c)$ covers a range whose relative magnitude increases with $\tau_c$.

The alternative way to interpret Figure~\ref{fig.R0_var} is as an
inverse problem.  In the inverse problem approach, a measurement 
of any given $\tilde{\tau}_c$ from a single trajectory
is compatible with a range of theoretically expected $\tau_c$ values.
For example, given measurement of $\tilde{\tau}_c=20$ days, then
 the associated CI of compatible $\tau_c$ 
is obtained by scanning horizontally (the red line)
for intersections with
the forward problem distributions (the black lines).
This intersection is estimated by linear interpolation of the 2.5\% and
97.5\% CI for $p(\tilde{\tau}_c|\tau_c)$ across different values of
$\tau_c$.  In this example, we estimate the CI of $\tau_c$ to be 16.6--27.9 days
given a single measurement of $\tilde{\tau}_c=20$ days. 

The CIs for $\tau_c$ from the case data is directly translated to CIs for
${\cal{R}}_0$ by utilizing a generating function approach \cite{wallinga_2007}. 
This method involves determining the moment generating function associated with the distribution of the age of secondary infections resulting from a single infectious individual in an otherwise susceptible population. This distribution varies with $\tau_c$ and specifies the initial deterministic dynamics. Converting the $\tau_c$ CI to ${\cal{R}}_0$ CI is shown on the right panel of Figure~\ref{fig.R0_var}. For the mock case data with $\hat{\tau}_c=20$ we obtain ${\cal{R}}_0 = 2.10$ with a CI of $1.80-2.45$.
In summary, these uncertainty ranges arise solely due to process noise
and were identified by quantifying COMETS in the SEIRD model framework. 

\subsection{Confidence intervals for ${\cal{R}}_0$ for the EVD outbreak
in W. Africa}
We now apply the hybrid stochastic-deterministic approach to estimate the confidence limits in ${\cal{R}}_0$ for early outbreak dynamics of EVD in Guinea, Liberia and Sierra Leone.
The present method differs from that of the previous section in one key way.
Previously, we varied rates of transmission to 
yield models with different theoretical $\tau_c$. Here,
we vary rates of transmission as well as the proportion of post-death transmission, $\rho_D$. Due to identifiability issues, measured values of $\hat{\tau}_c$ can correspond to different \Ro\ depending on the underlying parameters of the model~\cite{weitz_2015,eisenberg2015}. Hence, we systematically vary \Ro\ between ensembles rather than first varying $\tau_c$ and then transforming these into ranges of \Ro.
We construct the ensembles in a pseudo-Bayesian framework. Specifically, we account for uncertainty in the fraction of infections transmitted by the deceased by considering a uniform distribution of values of $\rho_D$ between $0.1-0.4$ (see ~\ref{sec.app.pseudo}).
In this way,
we account for uncertainty in the distribution of times to secondary transmissions in addition to uncertainty from process noise. Additionally, the total population for each country is set to match census data ($N = 4.3*10^6$ for Liberia, $N = 6.1*10^6$ for Sierra Leone, and $N = 11.8*10^6$ for Guinea).

Case data was obtained from the WHO \cite{eboladata}. 
We use cumulative counts and consider dynamics in each country given a
start date at which at least 50 cumulative infections have occurred and a final 
date of September 7, 2014. We choose this time period to reflect the onset of 
the epidemic across all three countries. 
The measured characteristic time, $\hat{\tau}_c$ is obtained by fitting an 
exponential to the CCC for
each country after the trigger population is reached. 
The lowest value of cumulative case 
counts above 50 for each country is also considered the trigger population for the 
stochastic simulations. We consider a SEIRD model with a gamma distributed 
exposed period with $n=2$ classes in accordance with previous analysis \cite{teamebola}.
The ensemble of simulated stochastic trajectories are conditioned on the epidemic remaining throughout the census period, as was the case with the EVD case data.
The fits to the stochastic trajectories are subject to
the same conditions as the case data. 

\begin{figure}[h]
 \includegraphics[width=0.3\textwidth]{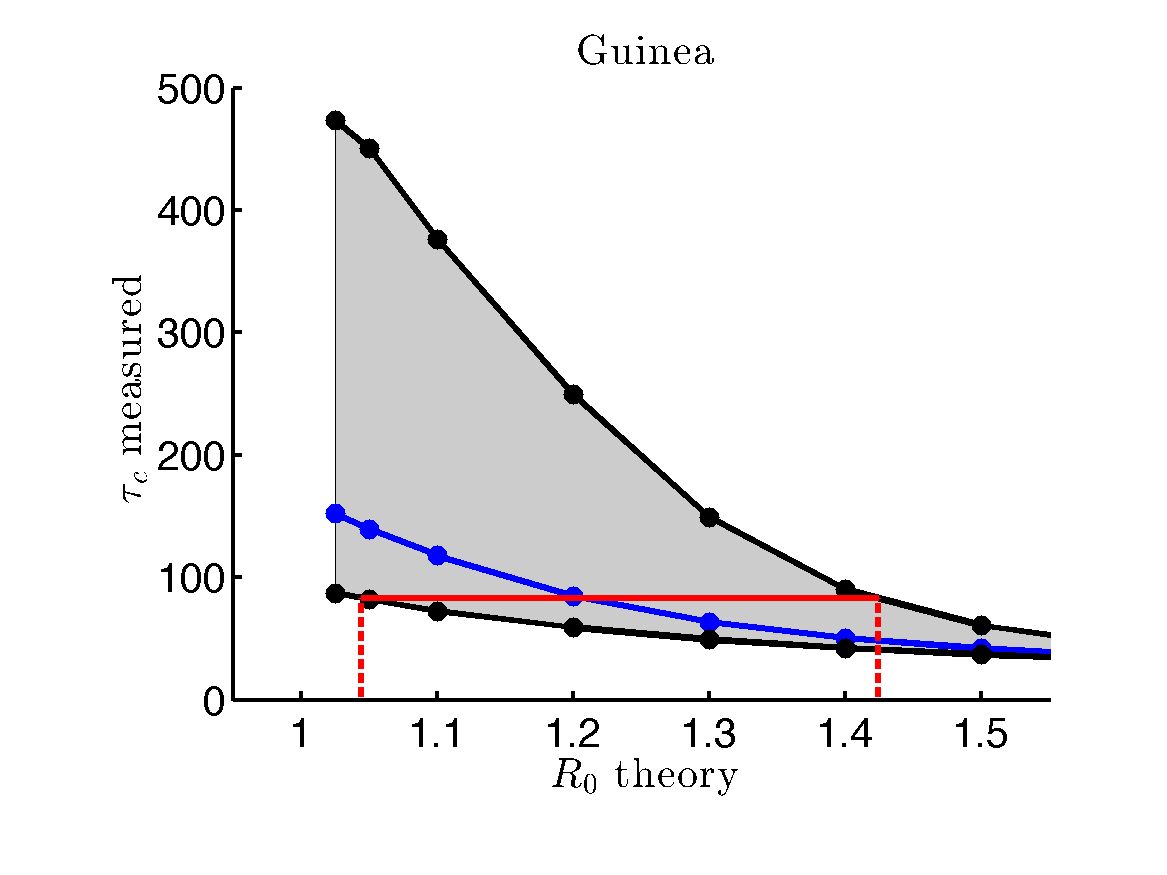}
  \includegraphics[width=0.3\textwidth]{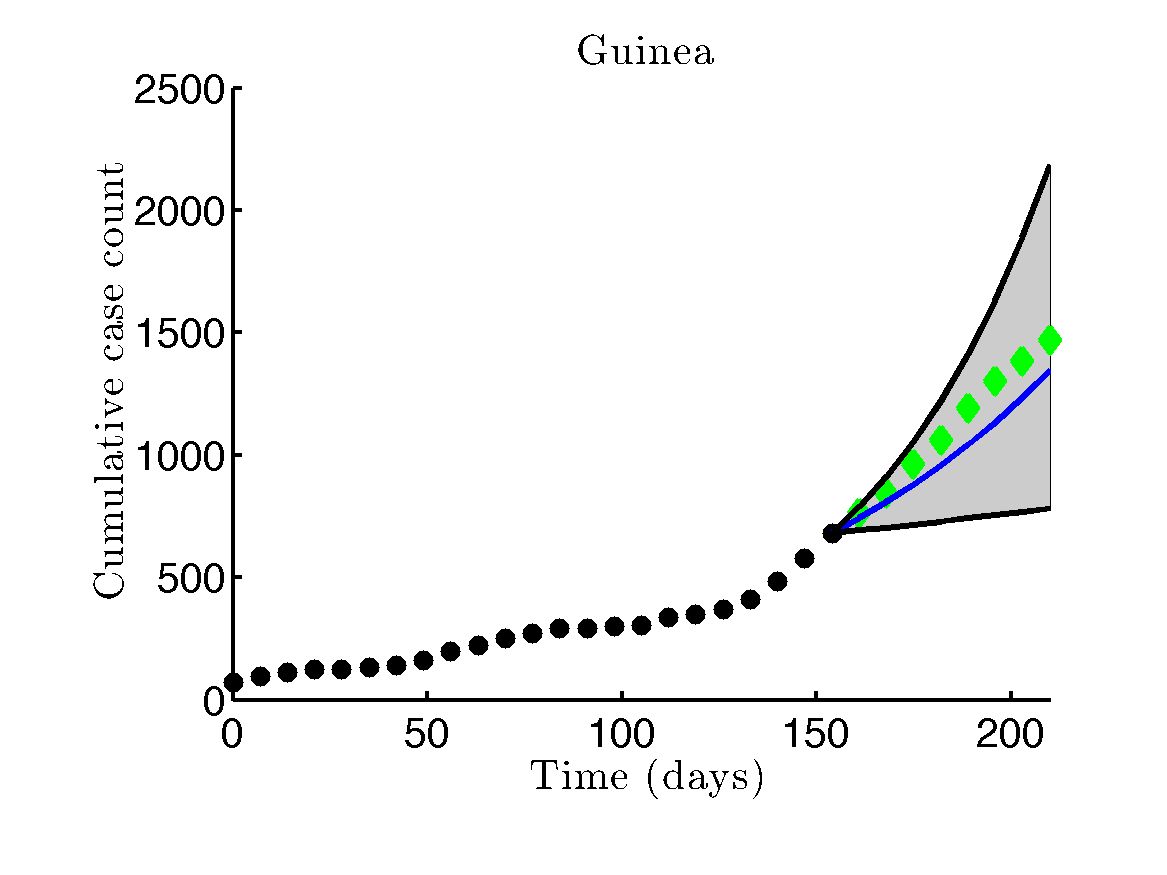}
    \includegraphics[width=0.3\textwidth]{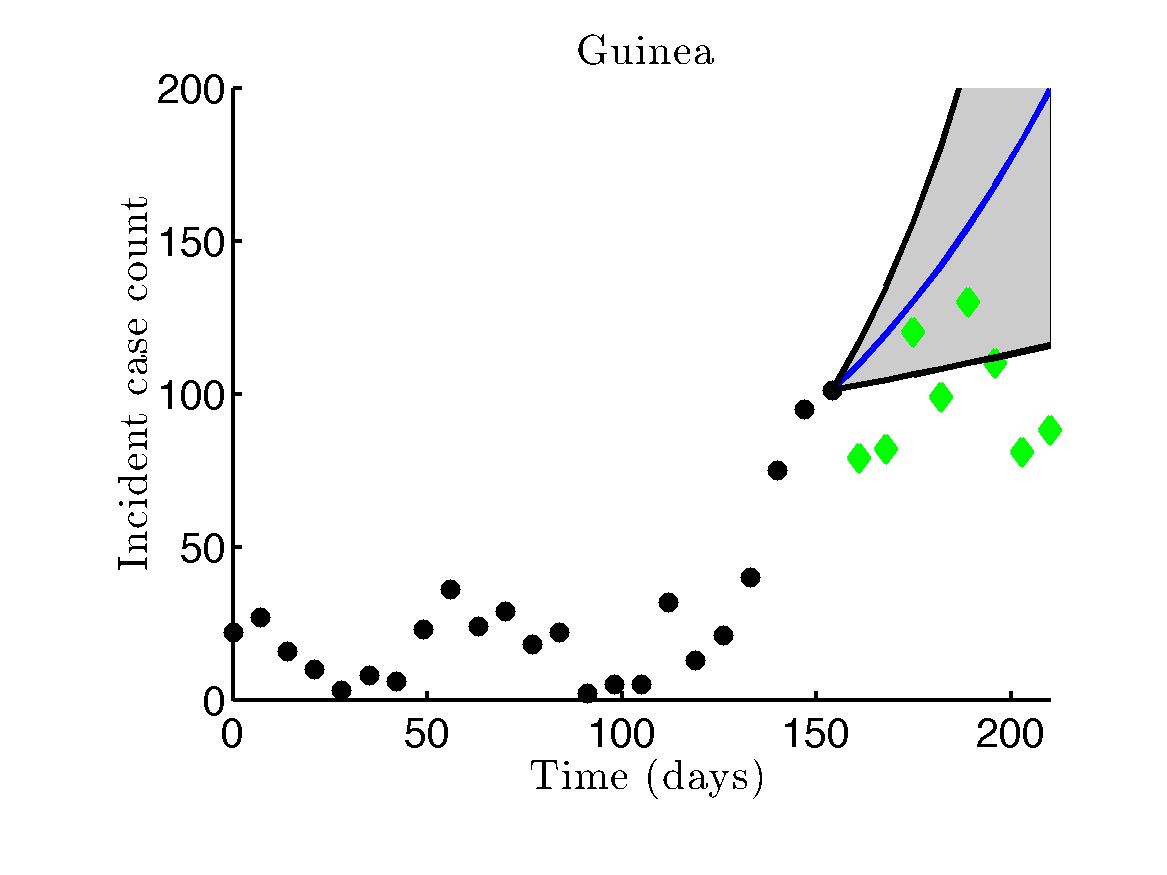}\\
 \includegraphics[width=0.3\textwidth]{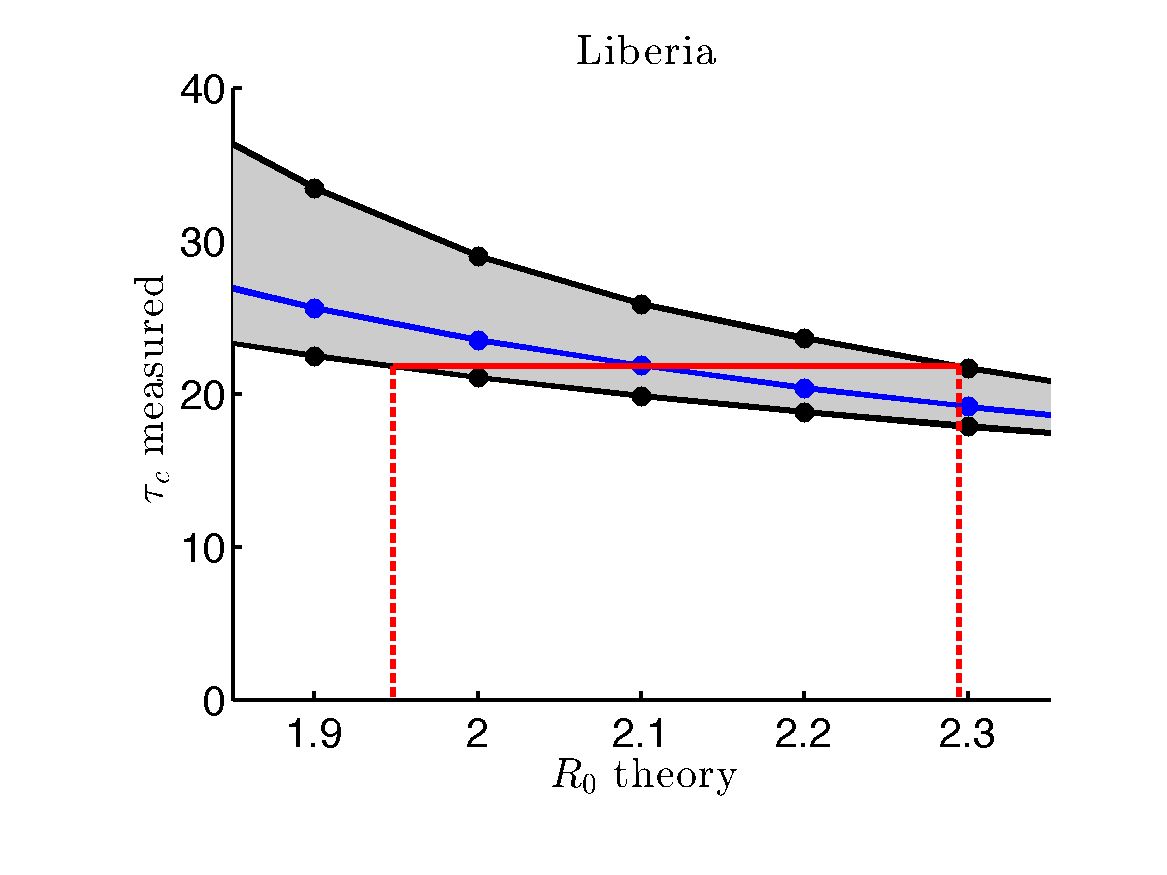}
 \includegraphics[width=0.3\textwidth]{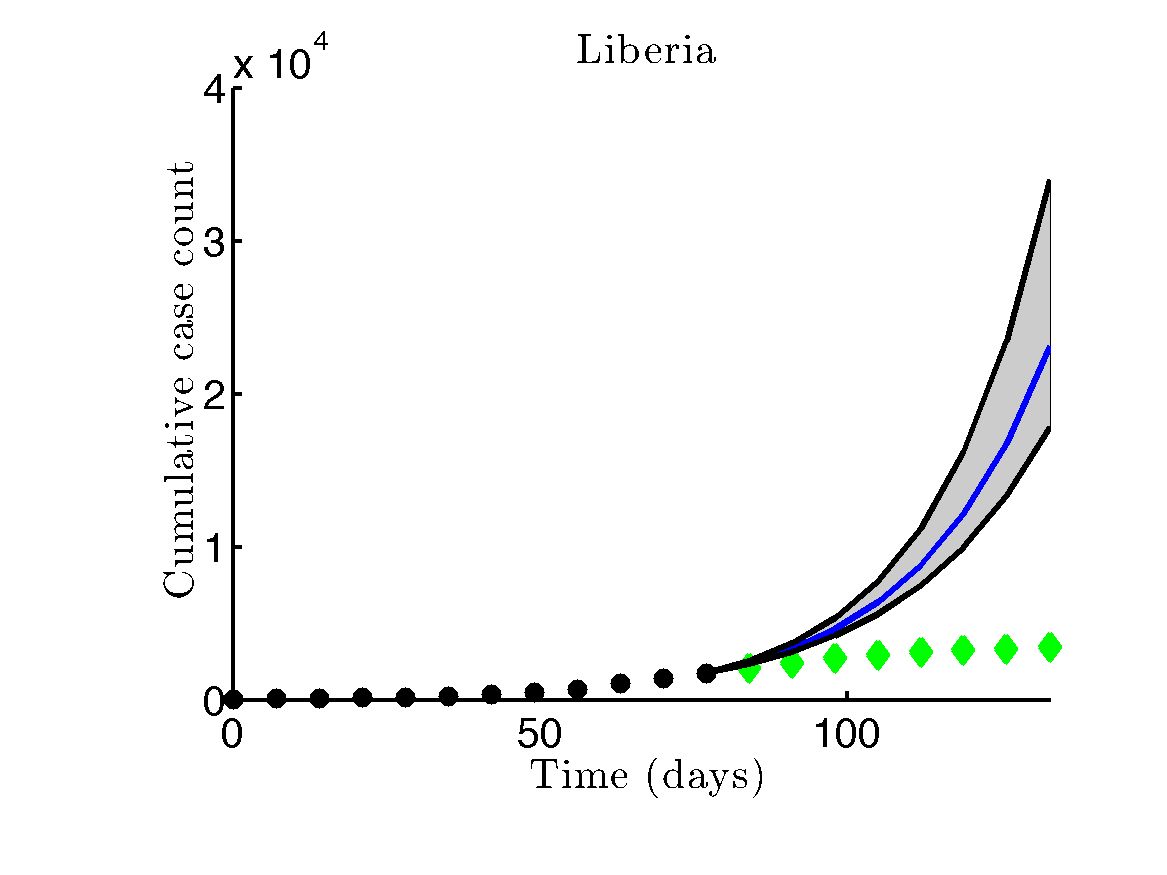}
     \includegraphics[width=0.3\textwidth]{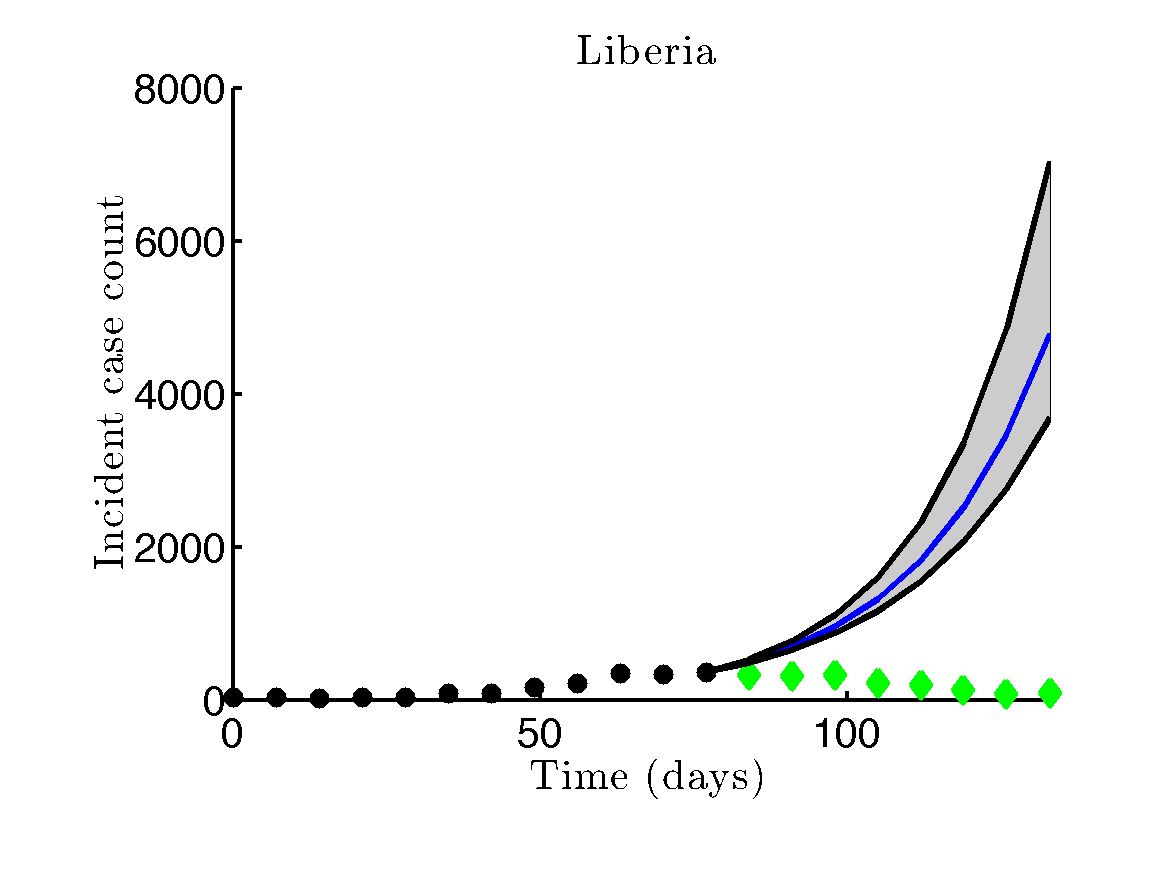}\\
 \includegraphics[width=0.3\textwidth]{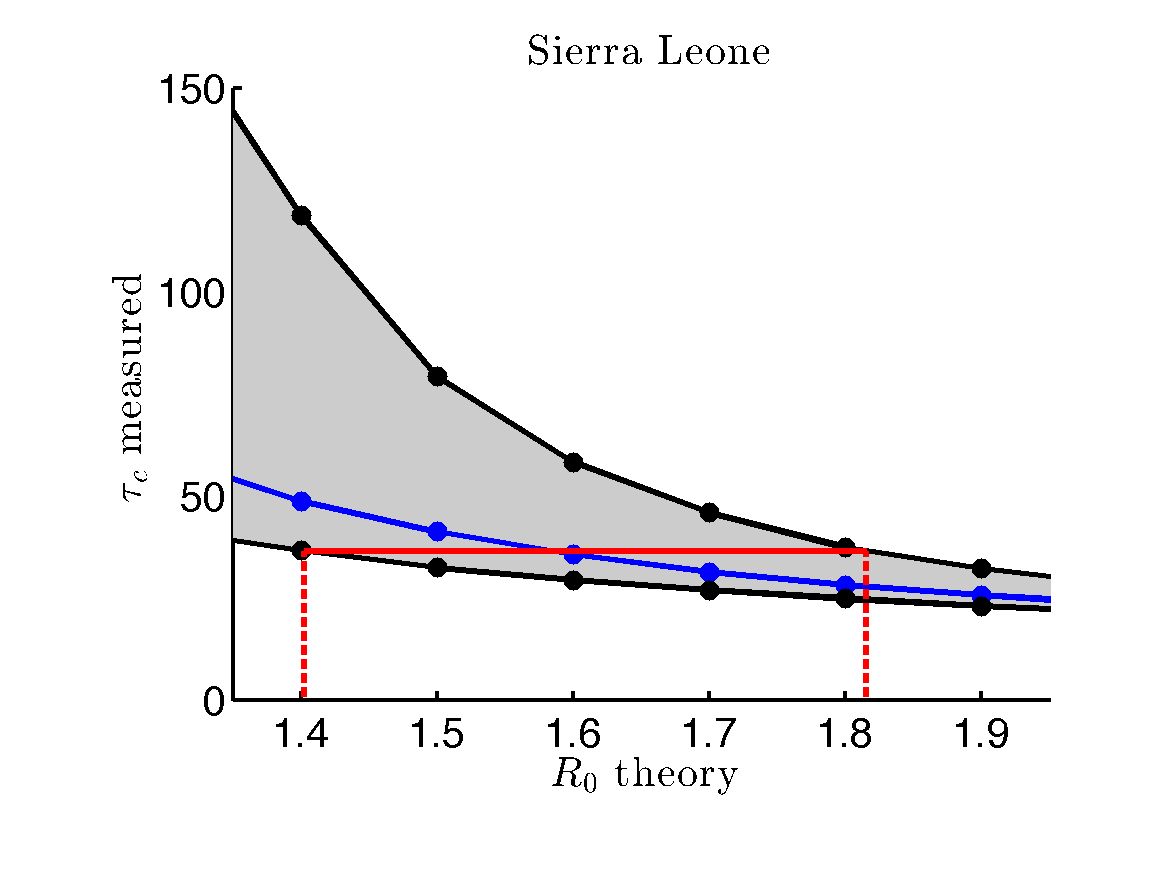}
 \includegraphics[width=0.3\textwidth]{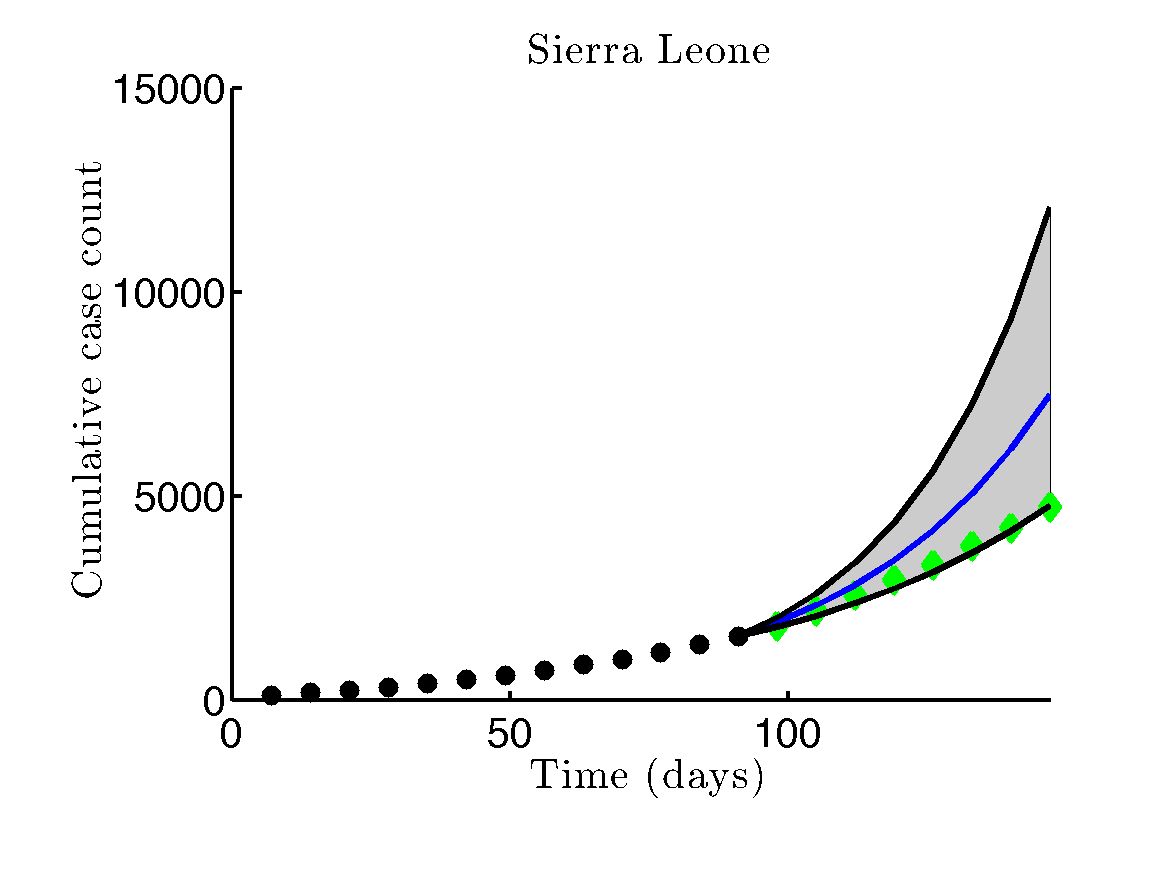}
     \includegraphics[width=0.3\textwidth]{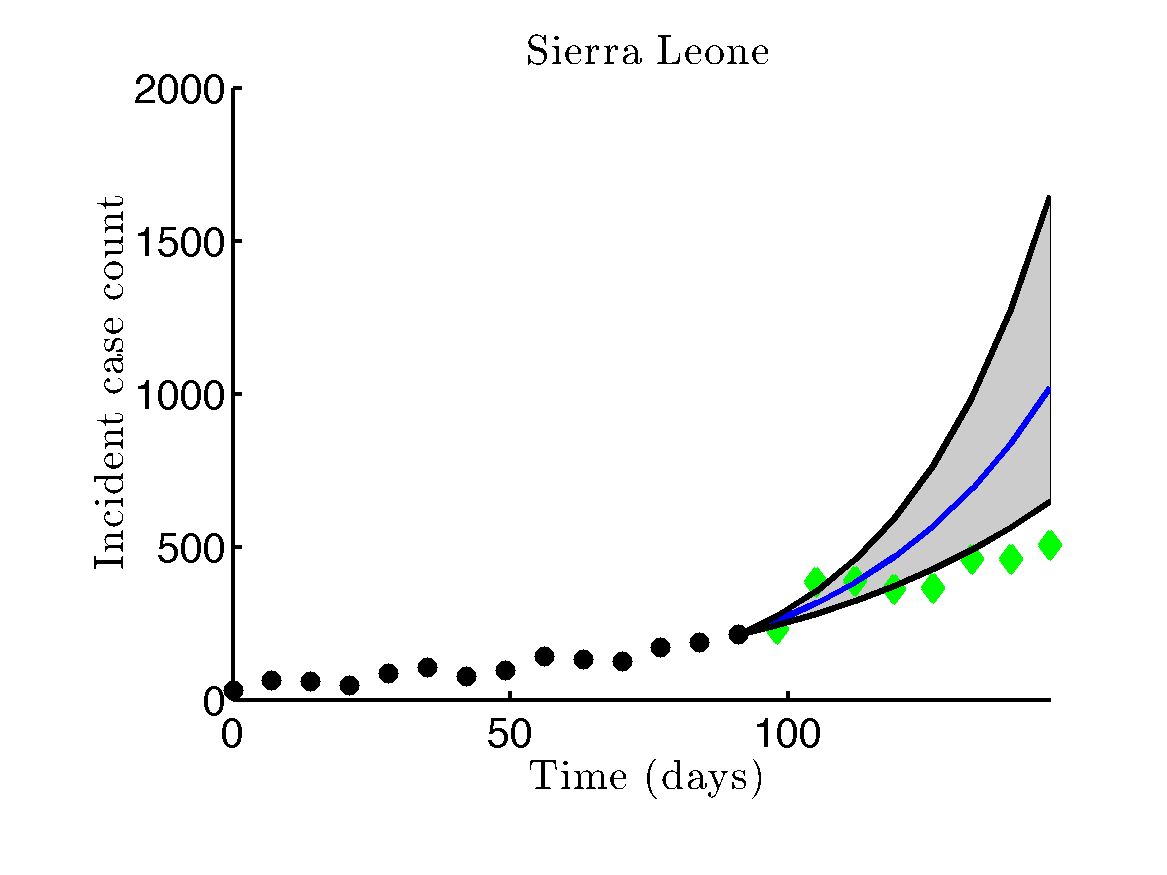}
      \caption{{\bf Estimated ${\cal{R}}_0$ CIs determined by the intersection of the measured case data $\hat{\tau}_c$ with the measured $\hat{\tau}_c$ CIs across a range of theoretical ${\cal{R}}_0$ for each country.} (Left column) Characteristic times estimated from case data are given by the height of the red line on the y-axis and the intervals are determined by the intersection of the red line with boundary of the CIs projected onto the x-axis. The blue line refers to the median of the distributions. (Top) Guinea, (Middle) Liberia, (Bottom) Sierra Leone. Table~\ref{tab:country_ci} shows the quantitative results for the CIs. (Middle Column) Cumulative case data with projections based on ${\cal{R}}_0$ CIs. The blue line corresponds to the estimate based fitting the case data. The gray area delineates the middle 95\% of the projection. The green diamonds are the reported cumulative case data. (Top) Guinea (Middle) Liberia and (Bottom) Sierra Leone. (Right Column) Incident case data with projections based on ${\cal{R}}_0$ CIs. The blue line corresponds to the estimate based fitting the case data. The gray area delineates the middle 95\% of the projection. The green diamonds are the reported incident case data. (Top) Guinea (Middle) Liberia and (Bottom) Sierra Leone.}
\label{fig:countryci}
\end{figure}
Resulting CIs of the characteristic time for Liberia, Sierra Leone, and Guinea are shown in the left column of Figure~\ref{fig:countryci}. The characteristic time measured from the case data specifies the location of the red line along the y-axis. For each value of theoretical ${\cal{R}}_0$ we obtain a distribution of measured $\tau_c$ from $10^4$ simulated stochastic trajectories. The confidence intervals for the reproductive number are determined by where the red line intersects the lower and upper limits of central
95\% characteristic time distribution for each theoretical ${\cal{R}}_0$. Country-specific estimates of ${\cal{R}}_0$ and associated
CIs are shown in Table ~\ref{tab:country_ci}. 
Again, these CIs represent the combined
uncertainty of estimating ${\cal{R}}_0$ from a single stochastic trajectory given
uncertainty in the distribution of times to secondary infection. 
Hence, they represent a lower bound of uncertainty in ${\cal{R}}_0$ given
additional uncertainty arising from observation noise.
These CIs are larger than many early estimates, e.g.,~\cite{althaus2014}.

In the middle (right) column of Figure~\ref{fig:countryci} we project forward the uncertainty in the cumulative (incident) case counts based on our CIs. We project forward by 8 weeks to show the range of expected case counts over time assuming no further control measures are implemented. The case data lies within the projection for Guinea and Sierra Leone, but are substantially lower for Liberia, likely because transmission parameters had already changed substantially, due to behavior change, control measures, or both, or due to structural
differences in the epidemic process~\cite{chowell_curr2014}.
Overall, we expect large uncertainty in projected case counts
during the exponential phase of the epidemic, due to process noise and the generation-interval uncertainty alone.

\begin{table}[h!]
\begin{tabular}{ | l | c | c |}
\hline
    \bf Country & \bf $\mathcal{R}_0$ & \bf $\mathcal{R}_0$ CI\\ 
    \hline
    Liberia & $2.06$ & $1.93-2.27$\\
    \hline
    Sierra Leone & $1.71$ & $1.40-1.82$\\
    \hline
    Guinea & $1.24$ & $1.04-1.42$\\
    \hline
\end{tabular}
\caption{\textbf{Confidence intervals of the SEIRD model given EVD case data in Liberia, Sierra Leone, and Guinea.} The ${\cal{R}}_0$ values are obtained from fitting a Poisson regression to the case data. The CIs are calculated based on Poisson fits to an ensembles of $10^4$ stochastic trajectories. The models used to obtain the trajectories varied in uniformly in $\rho_D$ between [.1, .4].}
\label{tab:country_ci}
\end{table}
\section{Discussion}
A stochastic implementation of an epidemic leads to significant variation in the realized time series of infectious
individuals, due to ``process noise''~\cite{keeling_2007} -- the stochastic sequence of discrete events in which individuals become infected, infect others, and eventually recover or die. The difference between trajectories predicted in deterministic vs.~stochastic models has been observed and studied for decades in other disease contexts~\cite{gibson_1998,ionides_2006,cauchemez_2008,ma_2014}. Here, we explored the practical consequences of such differences when trying to infer epidemiological parameters, including ${\cal{R}}_0$, at the early stages of an epidemic, and showed that stochastic variation constrains the extent to which confidence limits in ${\cal{R}}_0$ can be narrowed.  

In practice, time series that are used for fitting dynamical epidemiological models are drawn from the early stages of an epidemic. At such early stages, a single trajectory may be well fit by a single exponential rate from which ${\cal{R}}_0$ can be estimated.
Yet, the trajectories within an ensemble generated given the same underlying parameters will be fit by a distribution of exponential rates. Hence, the 95\% CIs for an estimate of ${\cal{R}}_0$ when using a deterministic model are significantly broadened due to process noise.

This general limitation of fitting deterministic models applies to the study of EVD. Multiple groups have proposed model variants of EVD dynamics, fit deterministic models to case data, and then used such fits to estimate ${\cal{R}}_0$, including associated CIs (e.g.,~\cite{althaus2014,lewnard_2014,pandey_2014,teamebola}).  The time range we used for fitting includes the period from onset as defined by the time with greater than 50 cumulative infections to early September 2014. This range spans from the onset of the epidemic to reported cumulative case counts of nearly 700 in Guinea and over 1500 in both
Sierra Leone and Liberia. Using a stochastic implementation of a SEIRD model, we inferred ${\cal{R}}_0$ to be 1.24 for Guinea (1.04-1.42 95\% CI), 2.06 for Liberia (1.93-2.27 95\% CI), and 1.71 for Sierra Leone (1.40-1.82 95\% CI).  
The estimates and ranges reflect three assumptions: (i) the model structure; (ii) the prior assumptions about the exposed and infectious periods, time to bury and probability of death; (iii) the availability and quality of epidemic case data. In particular, the second assumption involves varying the distribution of time to secondary infection. These CIs denote limits to inference in this class of model imposed by the nature of the data available: a single stochastic epidemic outbreak. The original estimates of CIs from~\cite{althaus2014} are evidently too narrow, yet even later estimates should be revisited using the hybrid approach proposed here.

The present method is complementary to alternative, profile-likelihood based approaches \cite{king_2015}. We obtain distributions for ${\cal{R}}_0$ 
by fitting to ensembles obtained by stochastically simulating a range of deterministic models. We marginalized our ensembles so that the trajectories in the ensembles were obtained from underlying models with the same theoretical $\tau_c$. Previous work used particle filtering techniques to obtain a distribution based on fitting to the data~\cite{king_2015}. This approach estimates uncertainty due to fitting deterministic models to a single, stochastic trajectory. 
In our case, we find that differences between fitting to cumulative or incident data are negligible, so long as the quality of individual fits within an ensemble is not used as the primary source of information on the CIs (see \ref{sec.app.ccc}).
We note that our methodology of estimating
COMETS is similar to the method of ``plausible parameter sets''
advocated for use in estimating disease-associated parameters during 
outbreaks\cite{drake2015}. Our goal is not to replace any previous methods but instead to provide an alternative flexible and computationally simple framework to estimate confidence intervals for an emerging epidemic.

The hybrid approach comes with certain precautions.  First off, this method is does not dictate the underlying model structure. Thus the relevance of the results relies on how accurately the actual dynamics of the epidemic are captured by the chosen underlying model. Of course, this is an issue whenever modeling actual epidemics and should be addressed prior to implementing our method.
Regardless of specific modeling choices, nonlinear models often generally display sloppiness such that many combinations of parameters
have little effect on certain system dynamics~\cite{gutenkunst_2007}.
The identifiability problem~\cite{weitz_2015,eisenberg2015} 
also applies here (see \ref{sec.app.identify})
, so that
the strength of fit does not necessarily exclude a range of
compatible mechanisms.
Implementing the current, hybrid approach should be
straightforward to adapt
to both well-mixed and spatially-explicit
models of disease transmission.
For example, 
a recent analysis of spatial data suggested that apparent exponential
dynamics is a result of aggregation of local epidemics~\cite{chowell_2014}.  
Yet, we expect that high-performance implementations of stochastic
models will be required
for spatially explicit projections of outbreak
sizes and associated CIs for ${\cal{R}}_0$
at the early stages of an epidemic~\cite{merler_2015}.  

In summary, we should have expected to have less confidence 
in estimates of CIs of
${\cal{R}}_0$ at the outset of a EVD epidemic given process noise.
We hope that the current method, similar in intent to that of King et al.~\cite{king_2015},
provides an accessible route for estimating realistic CIs for ${\cal{R}}_0$ 
in epidemics.  In practice, the 95\% confidence
intervals in ${\cal{R}}_0$ estimated from stochastic model fits
will be broader than that estimated from deterministic model
fits to cumulative case data.
Estimates of CIs using the current hybrid approach
represent a lower bound of uncertainty due to stochastic sources of noise.
As an epidemic continues and the number of infected individuals increases, observation noise contributes a relatively larger proportion of uncertainty as compared to process noise \cite{ma_2014}. 
Remaining realistic about the limits to
confidence in model fits should also be incorporated
into public health practice, e.g., 
when projecting the necessary scope of intervention
based on ``optimal'' fits~\cite{meltzer_2014}.
We encourage the academic, governmental, and non-governmental
public health
communities to consider 
incorporating unavoidable uncertainty into their decision making
pipelines when responding to emergent disease outbreaks.
%
%
%\section{References}
\bibliographystyle{siam}
\bibliography{cleanbib}

\appendix
\section{Appendix}
\subsection{Generating function approach to link characteristic times and
transmission rates}
\label{sec.app.generate}
The generalized SEIRD model includes $n_e$ number of exposed classes each
of duration $T_e/n_e$.  Given a value of $r_0=\tau_c^{-1}$ and
$\rho_D$, then the generating function approach of Wallinga
and Lipsitch~\cite{wallinga_2007,weitz_2015} can be used to derive
the following relationships:
\begin{eqnarray*}
{\cal{R}}_0 &= \frac{1+T_ir_0}{\left(\frac{n_e}{n_e+T_e r_0}\right)^{n_e} \left(1-\rho_D+\frac{\rho_D}{1+T_d r_0}\right)},\\
\beta_I &= \frac{(1+T_d r_0)(1+T_i r_0) (\rho_D-1)}{\left(\frac{n_e}{n_e + T_e r_0}\right)^{n_e} T_i(T_d r_0 (\rho_D-1)-1)},\\
\beta_D &= \frac{(1+T_d r_0)(1+T_i r_0) (\rho_D)}{\left(\frac{n_e}{n_e + T_e r_0}\right)^{n_e} f T_d(T_d r_0 (\rho_D-1)-1)}.
\end{eqnarray*}

\subsection{Stochastic simulations of epidemic outbreaks}
\label{sec.app.model}
Stochastic realizations of the SEIRD model are simulated 
using the Gillespie framework~\cite{gillespie77}, given the ``reaction'' events in the following table:

\begin{table}[h!]
\begin{tabular}{ | l | c | c |}
\hline
    \bf Process & \bf Reaction & \bf rate/probability\\ 
    \hline
    Pre-death infection & $S+I\rightarrow E+I$ & $r_1=\beta_I S\frac{I}{N}$\\
    \hline
    Post-death infection & $S+D\rightarrow E+D$ & $r_2=\beta_D I\frac{S}{N}$\\
    \hline
    Onset of infectiousness & $E\rightarrow I$ & $r_3=\sigma=1/T_E$\\
    \hline
    End of infectiousness (survival) & $I\rightarrow R$ & $r_4=(1-f)\gamma=(1-f)/T_I$\\
    \hline
    End of infectiousness (death) & $I\rightarrow D$ & $r_5= f\gamma=f/T_I$\\
    \hline
    Burial & $D\rightarrow B$ & $r_6=\rho=1/T_D$\\
    \hline
\end{tabular}
\caption{\textbf{Discrete stochastic reactions that transition the state of an individual}}
\label{tab:country_ci}
\end{table}
Processes transition individuals who are susceptible (S), exposed (E), infected (I), deceased (D), recovered (R), and buried (B).
The total population is fixed at $N=S+E+I+R+B$.
Epidemics are initiated with one infectious individual in an otherwise susceptible population. Mathematically, the initial state is, in the respective ordering, $\mathbf{y}=(N_0-1,0,1,0,0)$ 
at $t=0$.
The total rate of outbreak-associated events is $r_{tot}=\sum_{i=1}^6 r_i$.
The time until next event is determined randomly such that 
$\delta t\sim \frac{-\log{\chi}}{r_{tot}}$
where $\chi$ is a uniformly distributed number between 0 and 1. In this way,
the time between events follows an exponential distribution with rate $r_{tot}$.
Then, the probability of each event is $r_i/r_{tot}$.  After selecting
an event and updating
the discrete number of individuals, the reaction rates are recalculated 
and the process continues.
The same framework can be extended to include multiple $n$ subclasses
within the exposed class, to capture the peaked nature of
the exposed period (centered around 11 days for EVD).  
Trajectories are complete when the epidemic dies out because
there are no more infectious individuals. In the present context,
we are interested in those trajectories that do not die out before the end of the simulation time.

\begin{figure}[h]
\includegraphics[width=0.75\textwidth]{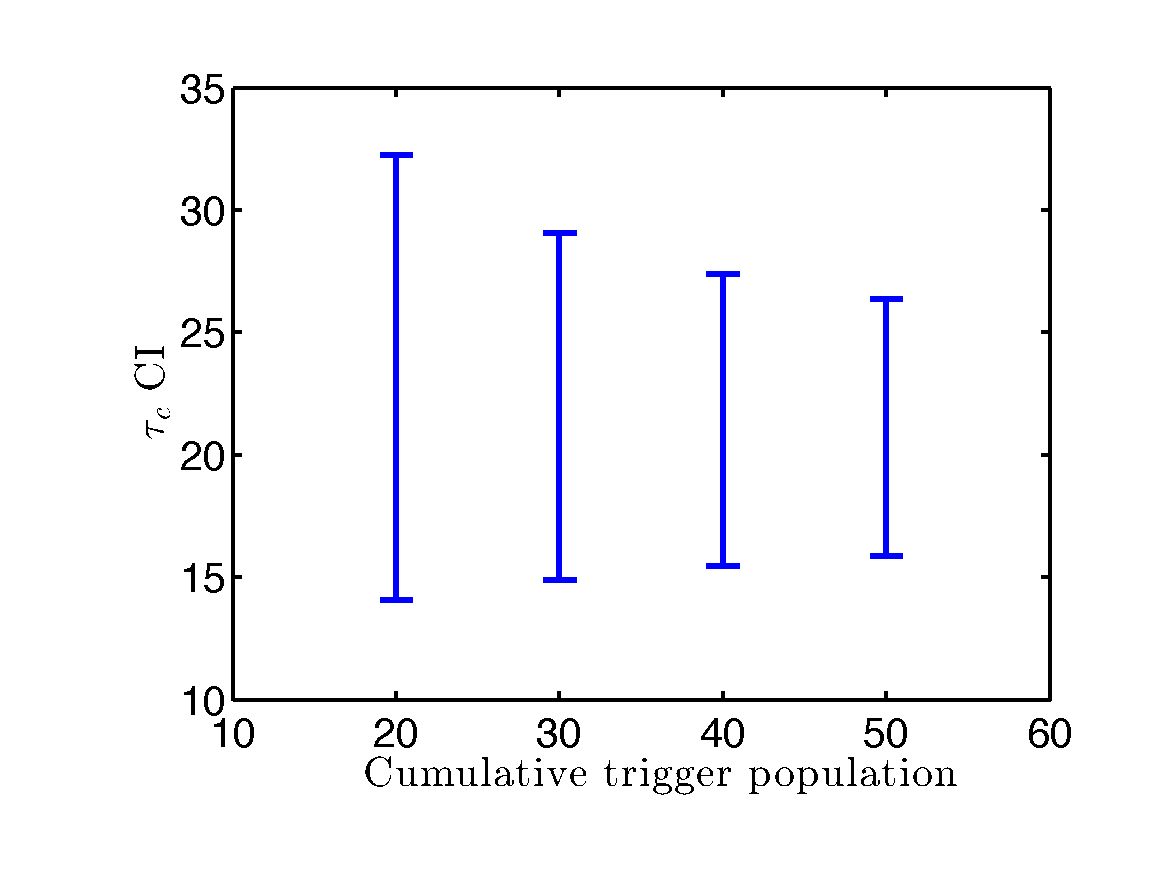}
\caption{{\bf The distribution of the characteristic time, $\tau_c$ increases with
decreasing trigger population.  }Ensembles of SEIRD stochastic dynamics
are simulated with parameters as in Figure \ref{fig.R0_var}. 
The ensemble includes $10^4$ trajectories with an epidemic that persists for 300 days.
In each case, estimates of $\tau_c$ are based on fits to the CCC for 
42 days after the trigger population is reached. 
\label{fig.Imin_compare}}
\end{figure}

\subsection{Pseudo-Bayesian approach for uncertainty in $\rho_D$}
\label{sec.app.pseudo}
The fraction of infections of EVD in west Africa due to transmissions from the deceased has been estimated to be between .01-0.3 \cite{teamebola}. However an estimate from a prior EVD outbreak in Congo is even higher \cite{legrand2007understanding}. To address this treat $\rho_D$ as uncertain with a uniform prior distribution between 0.1-0.4. We are interested with models that have a deterministic ${\cal{R}}_0$. For each ${\cal{R}}_0$, we randomly, uniformly sample  $\rho_D$ from the prior distribution. Each sample of ${\cal{R}}_0$ and $\rho_D$ determine the growth rate $r_0 = \tau_c^{-1}$ which in turn determines the infection parameters, $\beta_I$ and $\beta_D$. 

An ensemble of $10^4$ trajectories are obtained for each value ${\cal{R}}_0$ with $\rho_D$ sampled uniformly from [.1 .4] for each trajectory. The resulting distribution of measured characteristic times $\hat{\tau}_c$ can be interpreted as a marginal distribution across $\rho_D$. Even with a fixed ${\cal{R}}_0$, as $\rho_D$ varies the secondary infection distribution changes. Hence, the marginal distribution across $\rho_D$ can also be interpreted as a marginal distribution across the time to secondary infection distributions that all correspond to the same reproductive number of the disease in the population $\tau_c$.

\subsection{Comparing results from CCC and ICC data}
\label{sec.app.ccc}
\begin{figure}[h]
\includegraphics[width=0.75\textwidth]{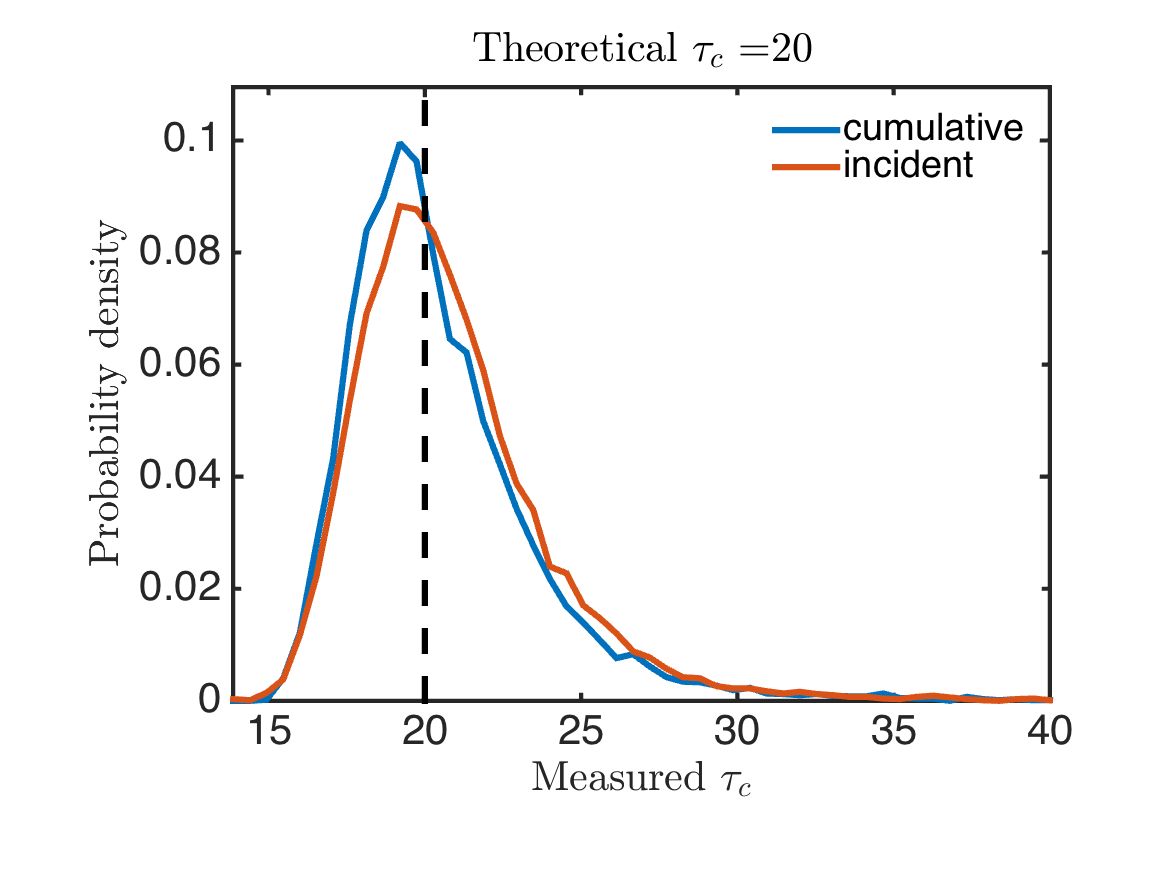}
\caption{{\bf Variation in regression-based fits of the characteristic time of case counts between estimates using cumulative or incidence
case count data. }The theoretical characteristic time of the underlying model is $\tau_c=20$ as shown by the black dashed vertical line. The ICC distribution has a median of 20.3 days with $95\%$ CIs of 16.6-28.5. The CCC distribution has a median of 19.9 days with $95\%$ CIs of 17.0-30.1.
\label{fig.ICC_CCC}}
\end{figure}

There are a number of potential pitfalls of using CCCs rather than ICCs.
Yet, King and colleagues in their critique of CCCs noted
that the summary statistics of the epidemic growth for SEIR models
were functionally equivalent when
fitted to an ensemble of CCCs and ICCs given the same underlying
disease parameters~\cite{king_2015}.  The difference, as they pointed out, was how to interpret
the \emph{quality of the fits} in inferring CIs.  Similarly, here we 
find that the resulting distributions, $p(\tilde{\tau}_c|\tau_c)$,
measured using either the CCC or ICC are similar but not equivalent
(Figure~\ref{fig.ICC_CCC}). Differences in the resulting CIs are minor, but the median of the ICC distribution is skewed larger than the theoretical value. These differences scale up to the overall analysis, but with minor effect on the $\tau_c$ and \Ro\ CIs. The resulting $\tau_c$ CIs from our synthetic data are 17.0--30.1 given fits to cumulative data and 16.6--28.5 given fits to incident data. The resulting CIs are approximately the same size.
This contrasts with prior results from profile likelihoods in which
the CIs as inferred from CCCs are contained within the CIs as inferred from ICCs \cite{king_2015}. 

\begin{figure}[h]
\includegraphics[width=0.75\textwidth]{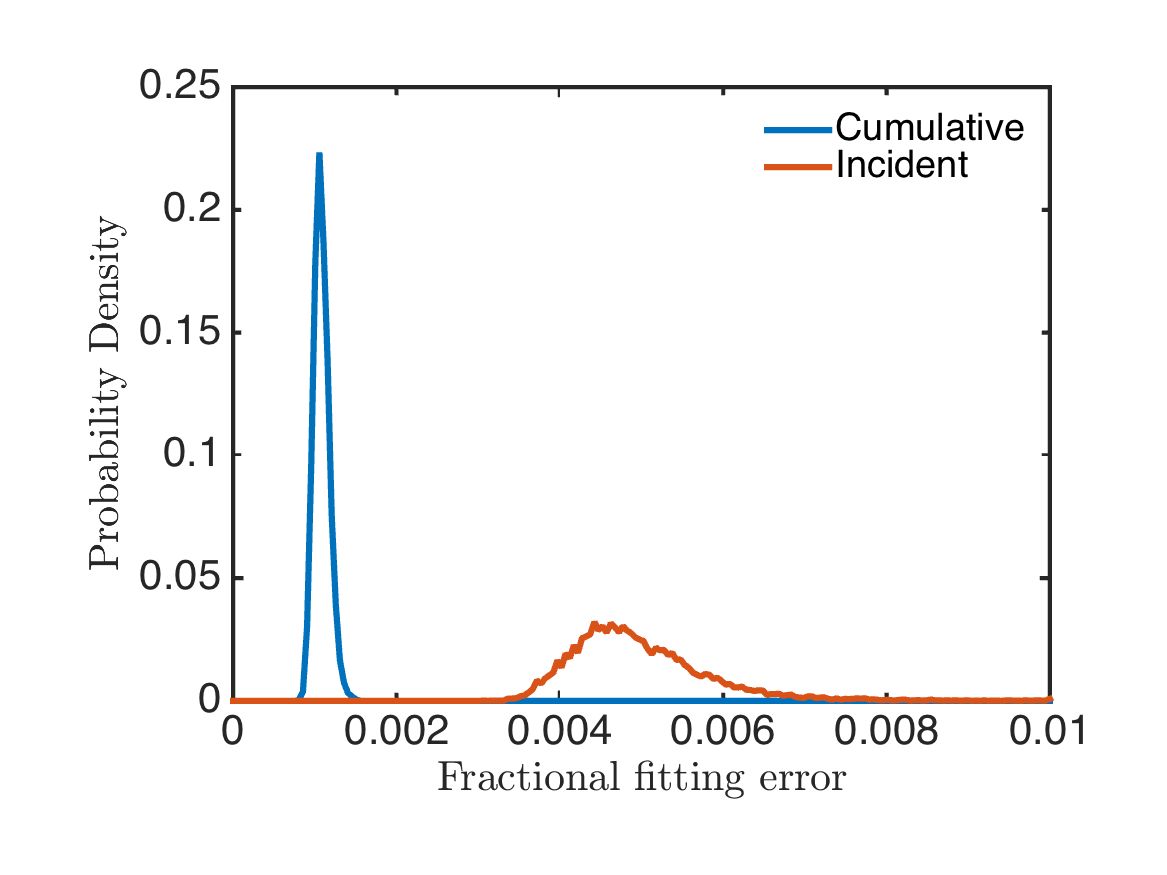}
\caption{{\bf Distributions of standard error of estimated growth rates, $r_0 = \hat{\tau}_c^{-1}$ for CCCs and ICCs arising from linearly fitting simulated data assuming deviations are Poisson distributed. }The underlying deterministic model for simulated data has a theoretical characteristic time of $\tau_c = 20$ days. 
\label{fig.fit_dist}}
\end{figure}

We emphasize that our method does not utilize the quality of any individual fit to generate the CIs for ${\cal{R}}_0$, for precisely the reasons cautioned by King and colleagues~\cite{king_2015}.
Estimating the growth rate, $r_0$, from linear regression is uncertain due to the fitting procedure itself. Associated with each estimate of $r_0$ are confidence intervals. We compare the distributions of errors due to fitting associated with the CCCs and ICCs of our simulated data in Figure~\ref{fig.fit_dist}. The errors associated with ICCs are larger than those associated with CCCs. It remains an open question as to whether/when
CCCs rather than ICCs are preferable when 
leveraging regression fitting to estimate $r_0$ given process noise
alone, rather than observation noise.

\subsection{Identifiability problem persists when inferring
relative fraction of post-death transmission from stochastic trajectories}
\label{sec.app.identify}
\begin{figure}[h]
\includegraphics[width=0.75\textwidth]{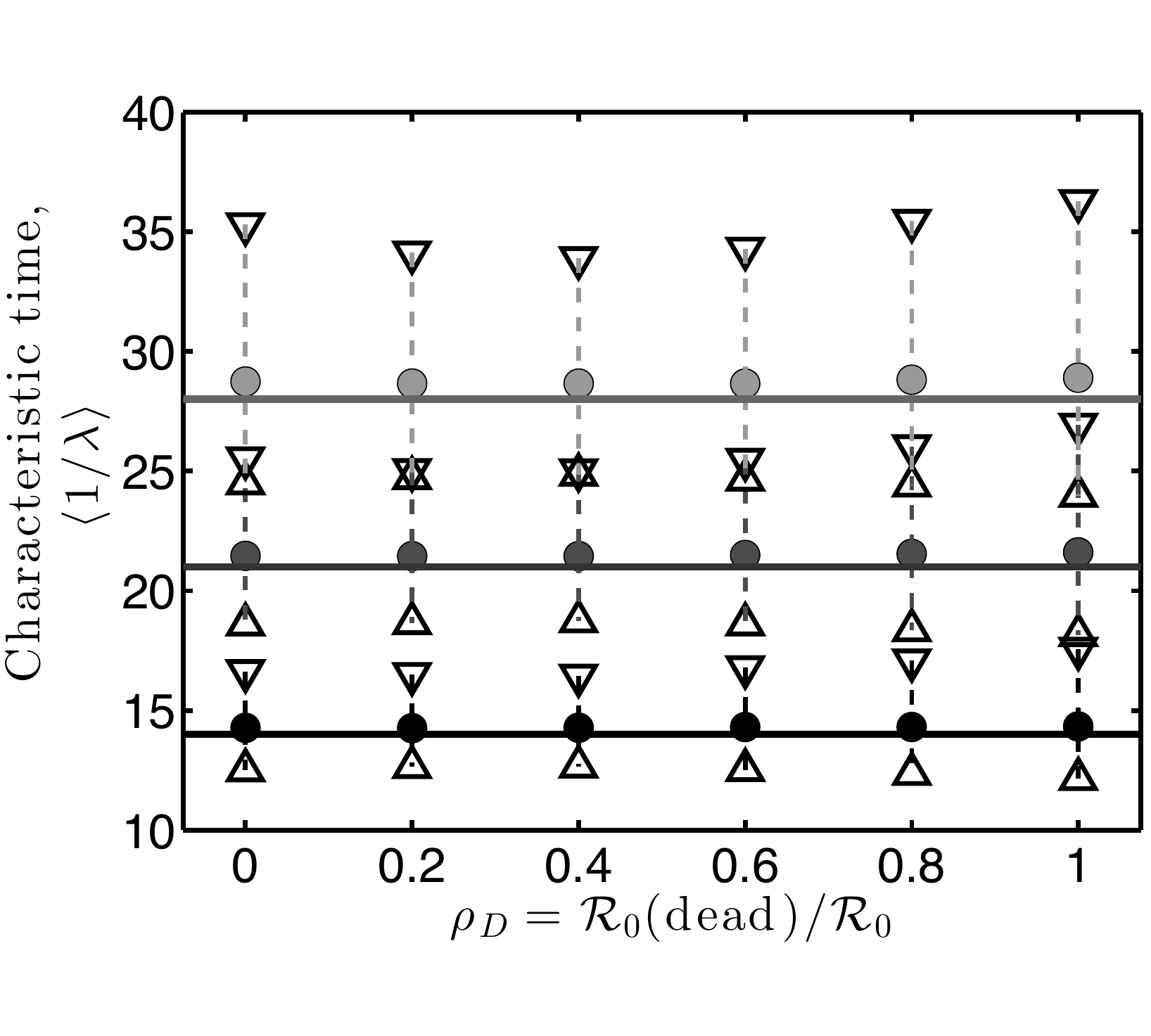}
\caption{{\bf Identifiability problem persists when fitting SEIRD models
to stochastic data.}The three scenarios correspond to cases where
the characteristic time, $1/r_0$=14, 21 and 28 days.  The realized
epidemic growth rates of stochastic trajectories are measured given
variation in $\rho_D$ from 0 to 1 in increments of 0.1.  Circles
denote the median characteristic time while triangles denote the
95\% confidence intervals from an ensemble of $10^3$ simulations per
condition.
\label{fig.lambda_var}}
\end{figure}

The measured growth rate of 
an epidemic, $\hat{r}_0=\hat{\tau}_c^{-1}$ can be used to infer the basic
reproductive number, ${\cal{R}}_0$. Using a generating function
approach, it can be shown that ${\cal{R}}_0=1/M(-r_0)$
where $M(z)=\int_0^\infty e^{za}g(a)\d a$~\cite{wallinga_2007}
where $g(a)$ is the normalized fraction of all secondary cases caused
by an infectious individual at ``age'' $a$ since infection.
A range of values of 
${\cal{R}}_0$ may be compatible with a single estimate of $\hat{r}_0$~\cite{weitz_2015}.
This uncertainty is a consequence of an identifiability problem
given uncertainty in the relative fraction of transmission
events that could be attributed to post-death transmission.
Here we ask: what is the variation in the 
growth rate, $r_0$, and characteristic time, $1/r_0$,
compatible with varying fraction of post-death transmission, $0<\rho_D<1$.
Figure~\ref{fig.lambda_var} shows the measured variation in 
the characteristic time, $\langle 1/\hat{r}_0 \rangle$ for
5000 ensembles for three different expected growth rates
$r_0=1/14$, 1/21 and 1/28 days$^{-1}$.  In each case, we varied
$\rho_D$ from 0 to 1 in increments of 0.2.  
As in the prior section, we find that the
characteristic time of an epidemic can vary substantially
for a fixed value of $r_0$.  Here, we also expect
that a range of mechanistic models can all yield the same
expected characteristic time.  
As is evident,
the identifiability problem highlighted in prior analyses
of deterministic models~\cite{weitz_2015,eisenberg2015} also applies in
the case of stochastic models.
For SEIRD models, the expected value
of the basic reproductive number increases with $\rho_D$.
Therefore, efforts to constrain
estimates of ${\cal{R}}_0$ from EVD case data will
be subject to inherent variability due, in part, to uncertainty
in mechanism, e.g., the relative fraction of post-death transmission,
and process, i.e., stochastic outbreak dynamics.

\section{Acknowledgments}
The work was funded by a grant to JSW from the Burroughs Wellcome Fund and from the Army Research Office grant \#W911NF-14-1-0402. We would like to acknowledge Luis F. Jover for reviewing the manuscript.

\end{document}